\documentclass[fleqn,usenatbib]{mnras}

\usepackage{newtxtext,newtxmath}

\usepackage[T1]{fontenc}
\usepackage{ae,aecompl}

\usepackage{graphicx}	
\usepackage{amsmath}	
\usepackage{amssymb}	
\usepackage{hyperref}
\usepackage{caption}
\usepackage{subfigure}
\usepackage{ulem}
\usepackage{color}
\usepackage[colorinlistoftodos]{todonotes}

\newcommand{\kpc}{\,\mathrm{kpc}}
\newcommand{\magt}{\,\mathrm{mag}}

\newcommand{\Gyr}{\, \mathrm{Gyr}}

\newcommand{\smp}{SkyMapper}

\newcommand{\masyr}{\,\mathrm{mas \,yr^{-1}}}
\newcommand{\kms}{\,\mathrm{km \,s^{-1}}}

\def\msun{{\rm\,M_\odot}}

\title[\smp\ view of the LMC]{A \smp\ view of the Large Magellanic Cloud: \\ The dynamics of stellar populations}

\author[Z. Wan et al]{Zhen Wan$^{1}$\thanks{E-mail: zwan3791@uni.sydney.edu.au},
Magda Guglielmo$^{1}$,
Geraint F. Lewis$^{1}$,
Dougal Mackey$^{2}$
\newauthor
and Rodrigo A. Ibata$^{3}$
\\\\
$^{1}$Sydney Institute for Astronomy, School of Physics A28, The University of Sydney, NSW, 2006, Australia\\
$^{2}$Research School of Astronomy \& Astrophysics, Australian National University, Canberra, ACT 2611, Australia\\
$^{3}$Observatoire Astronomique, Universit\'e de Strasbourg, CNRS, 11, rue de l'Universit\'e, F-67000 Strasbourg, France\\
}
\date{Accepted XXX. Received YYY; in original form ZZZ}

\pubyear{2019}

\usepackage{etoolbox}
\makeatletter
\patchcmd\@combinedblfloats{\box\@outputbox}{\unvbox\@outputbox}{}{%
   \errmessage{\noexpand\@combinedblfloats could not be patched}%
}%
 \makeatother
\begin{document}
\label{firstpage}
\pagerange{\pageref{firstpage}--\pageref{lastpage}}
\maketitle

\begin{abstract}
We present the first \smp\ stellar population analysis of the Large Magellanic Cloud (hereafter LMC),including the identification of 3578 candidate Carbon Stars through their extremely red $g-r$ colours. 
Coupled with {\it Gaia} astrometry, we analyse the distribution and kinematics of this Carbon Star population, finding the LMC to be centred at $(R.A., Dec.) = (80.90^{\circ}\pm{0.29}, -68.74^{\circ}\pm{0.12})$, with a bulk proper motion of $(\mu_{\alpha},\mu_{\delta}) = (1.878\pm0.007,0.293\pm0.018) \masyr$ and
a disk inclination of $i = 25.6^{\circ}\pm1.1$ at position angle $\theta = 135.6^{\circ}\pm 3.3^{\circ}$. We complement this study with the identification and analysis of additional stellar populations, finding that the dynamical centre for Red Giant Branch (RGB) stars is similar to that seen for the Carbon Stars, whereas for young stars the dynamical centre is significantly offset from the older populations. This potentially indicates that the young stars were formed as a consequence of a strong tidal interaction, probably with the Small Magellanic Cloud (SMC). 
In terms of internal dynamics, the tangential velocity profile increases linearly within $\sim3\ \kpc$, after which it maintains an approximately constant value of $V_{rot} = 83.6\pm 1.7 \kms$ until $\sim7 \kpc$. 
With an asymmetric drift correction, we estimate the mass within $7\kpc$ to be $M_{\rm LMC}(<7\kpc) = (2.5\pm0.1)\times10^{10}\msun$ and within the tidal radius ($\sim 30\ \kpc$) to be $M_{\rm LMC}(<30\kpc) = (1.06 \pm 0.32)\times10^{11}\ \msun$, consistent with other recent measurements.
\end{abstract}



\section{Introduction}

The Large Magellanic Cloud (LMC) is amongst the largest dwarf galaxies within the Local Volume \citep[see][]{2012AJ....144....4M}, and its complex evolutionary history is encoded in its present structure and dynamics.
As such, kinematic observations of the various components of the LMC have revealed the orientation and morphology of its stellar disk \citep[e.g.][]{Freeman1983,Meatheringham1988,VanderMarel2002,Olsen2011}, with a star formation history which has peaked at several points over the past $5\Gyr$ \citep{Harris2009}. 
Additionally, radio observations of the gaseous components of the LMC, and the more extensive Magellanic System, have revealed the signatures of historical LMC-SMC interactions \citep[e.g.][]{Staveley-Smith2003,Bruns2005,Tepper-Garcia2019}, events that are also now known to be encoded in the structure of the peripheral stellar component \citep{Olsen2002,Belokurov2016,Besla2016,Mackey2017,Choi2018,Nidever2018,Mackey2018,Vasiliev2018}. Most recently, \citet{Belokurov2019} identified low-surface brightness stellar arms around the LMC from the panoramic view of RGB stars, further highlighting the results of the LMC-SMC-MW interactions.

\begin{figure}
    \includegraphics[width=\columnwidth]{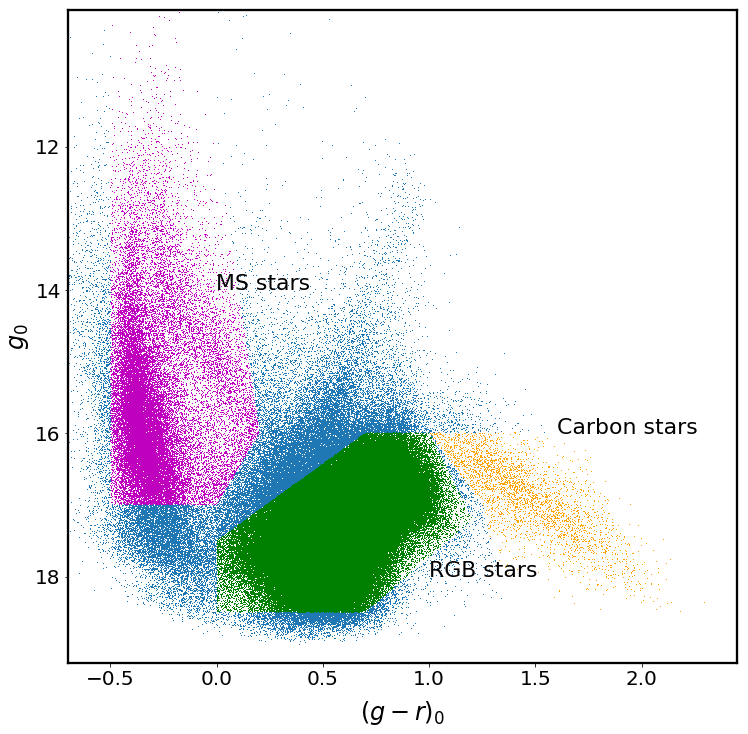}
        \caption{The SkyMapper $(g-r)_0$ vs $g_0$ CMD of stars within $10^{\circ}$ of $(R.A., Dec.) = (81^{\circ}.91,-69^{\circ}.87)$. Foreground stars with parallax measurements $\varpi > 0.1$ are excluded to reduce contamination. Young main sequence stars are dominant between $-1 < (g-r)_0 < 0$ (selected as the magenta region), and older evolved stars (RGB) are in $0 < (g-r)_0 < 1$ (green selected region). The reddest branch (orange points) is the Carbon Star population.} 
        \label{Fig:smp_cmd}
\end{figure}


The dynamical evolution of the LMC depends upon its mass. For example, \citet{Besla2007,Besla2010,Besla2012} proposed that the Magellanic system is currently on its first orbital pass around the Milky Way, requiring a total mass of $M > 10^{11} \msun$ \citep{Kallivayalil2013}. Moreover, recent simulation from \citet{Erkal2018} estimates the LMC mass to be $1.38\times10^{11}\ \msun$ from the perturbation on the Milky Way stellar stream. However, early observational measurements based upon internal kinematics find masses substantially smaller than this;  e.g. \citet{Meatheringham1988} estimate a mass of $6\times10^{9}\ \msun$ from planetary nebulae, whilst \citet{Kim1998} use H{\small\uppercase\expandafter{\romannumeral1}} dynamics to estimate the mass of LMC within $4\ \kpc$ to be $3.5\times10^{9}\ \msun$. This low mass LMC was rejected by \citet{VanderMarel2002} with stellar radial velocity measurements, although mass estimates are limted by the paucity of kinematic tracers at large radius.

\begin{figure*}
    \includegraphics[width=\textwidth]{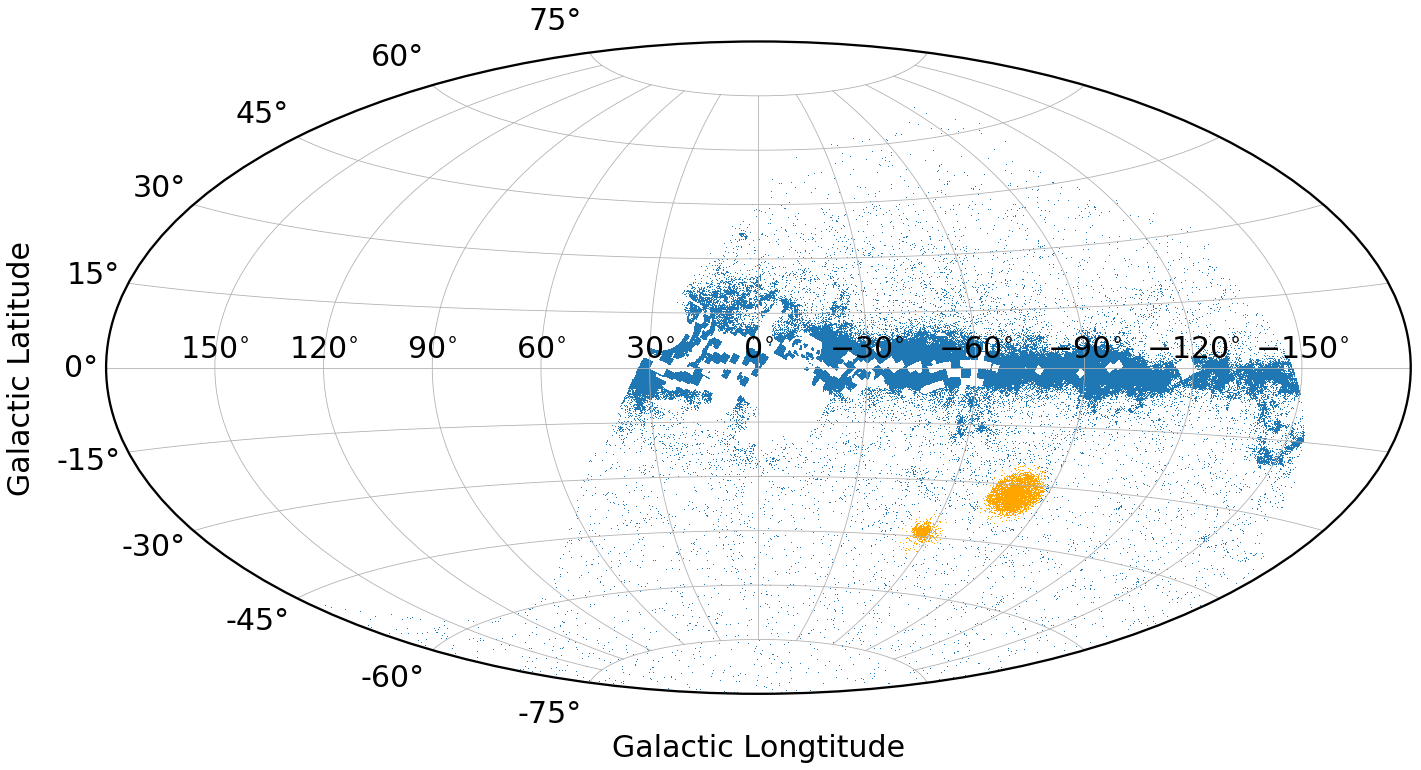}
        \caption{The distribution of stars with $(g-r)_0 > 1.2$ from \smp, with stars concentrated in the Galactic disk and Magellanic Clouds. The orange points denote the concentration of Carbon Star candidates within $10^{\circ}$ of $(R.A., Dec.) = (81^{\circ}.91,-69^{\circ}.87)$, the LMC centre from \citet{VanderMarel2002}, and within $4^{\circ}$ of $(R.A., Dec.) = (16^{\circ}.25,-72^{\circ}.42)$, the SMC centre from \citet{Stanimirovi2004}.
        }
        \label{Fig:carbon_sky}
\end{figure*}

\begin{figure}
    \includegraphics[width=\columnwidth]{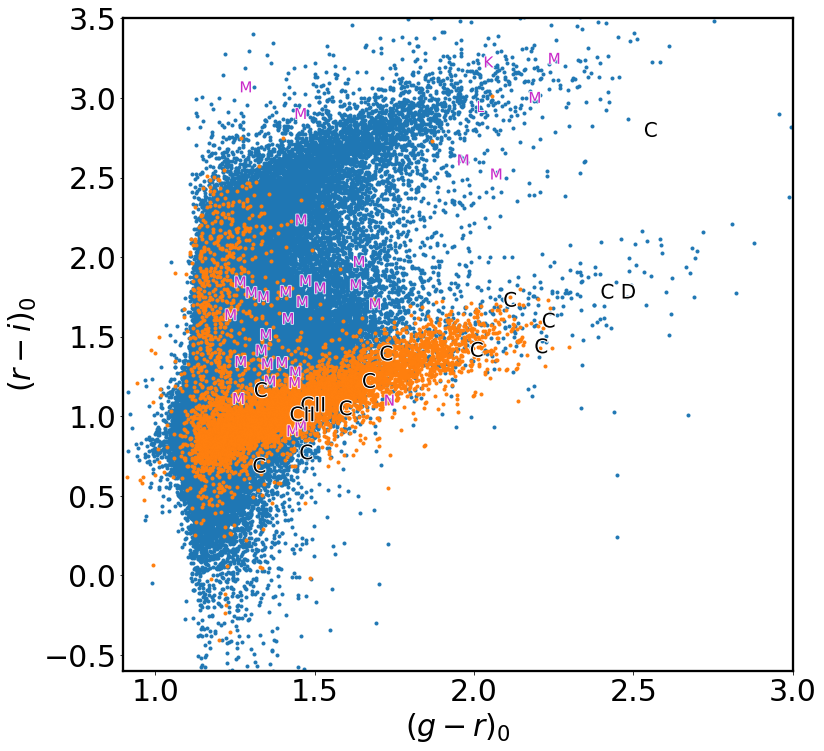}
        \caption{The colour-colour diagram of stars with $(g-r)_0 > 1$ from \smp. Orange points correspond to the stars in Magellanic Clouds region. Blue points are from other regions excluding the Magellanic Clouds. The text in this figure marks the spectral type and indicates the location in the two-colour plane of stars of different spectral types derived using spectra from the X-Shooter Spectral Library integrated over the \smp\ transmission curves (Sec.~\ref{sec:data}). Spectra marked with "C" indicate Carbon Stars from this library.}
        \label{Fig:gr_17} 
\end{figure}

\begin{figure}
    \includegraphics[width=\columnwidth]{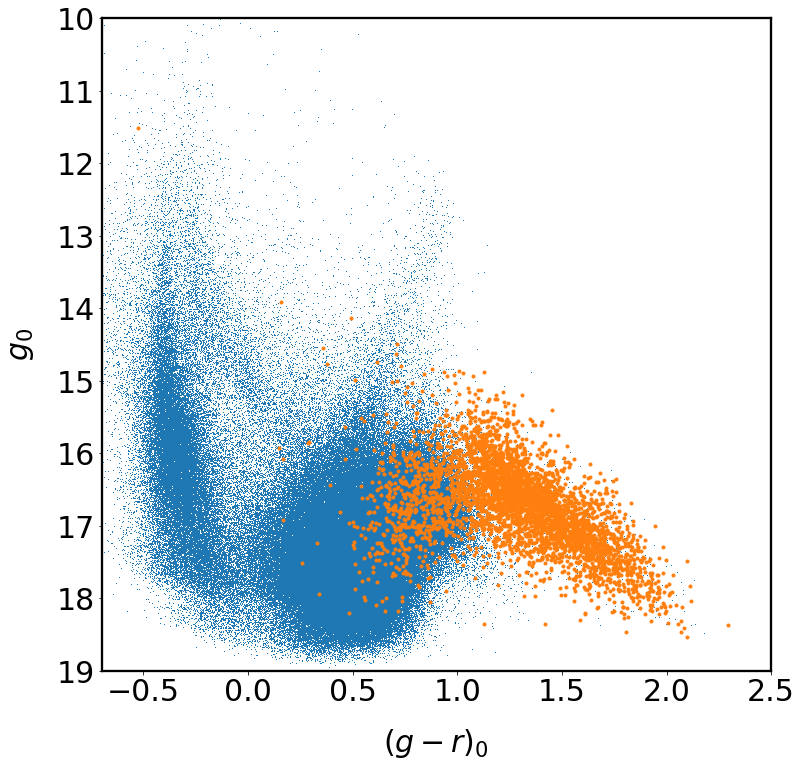}
        \caption{The \smp\ CMD of Carbon Stars from \citet{Kontizas2001} (orange points). We find a good match between their catalogue and our Carbon Star sample. This demonstrates  that the Carbon Stars are, in general, well separated from RGB stars in the \smp\ $g$ and $r$ bands, providing an effective way to identify and isolate these stars.}
        \label{Fig:matched_carbon}
\end{figure}

The dynamical interactions of the Magellanic Clouds can imprint differing signatures on different stellar populations, identifiable in global structure and phase space distributions. 
For example, the younger stellar population is observed to be more clumpy than older stars in both LMC and SMC \citep[e.g.][]{Zaritsky2000,Cioni2000b,Nikolaev2000,Belcheva2011,Moretti2014,Mackey2017}, and some carbon-rich AGB stars are likely to form a second (or third) disk  in the LMC \citep[e.g.][]{Graff2000,Olsen2011}. We can also see difference in the inclination and position angle estimations of the different disk populations \citep[e.g.][]{Kim1998,VanderMarel2001,Haschke2012,Subramanian2013,Deb2014,Subramanian2015,Jacyszyn-Dobrzeniecka2016,Inno2016}. By comparing different population, hence we can infer the history of the LMC.

The current \smp\ \citep{Wolf2018} and {\it Gaia} \citep{Gaia2} surveys provide a means of identifying different populations, especially Carbon Stars in the Magellanic Clouds,  with essentially minimal contamination, thus facilitating a detailed kinematic portrait of the LMC and enabling comparisons of different populations within same context. As detailed in the following sections, the \smp\ photometric system is ideally suited to identifying the Carbon Star population of the LMC; as luminous intermediate-age stars distinguishable by their broad carbon absorption, they represent excellent tracers of the structure and kinematics of the LMC. For instance, Carbon Stars have been used to measure the configuration of LMC (\citet[e.g.][]{VanderMarel2002} with catalogues from \citet{Kunkel1997,Hardy2001}.

In this contribution, we present the first results of our stellar population survey of the LMC, using the derived kinematics to determine its mass and compare the dynamical signatures of differing populations. In Sec.~\ref{sec:data}, we discuss the selection of different populations from the \smp\ derived colour-magnitude diagram (CMD) and describe their basic properties. In Sec.~\ref{sec:result}, we examine the derived kinematic profile using our Carbon Star sample and estimate the mass of the LMC. Additionally, we further present complementary analyses for young  and RGB stars. We conclude the paper in Sec.~\ref{sec:conclusion}.

\section{Data}
\label{sec:data}

The aim of \smp\ is to create a deep, multi-epoch, multi-colour digital survey of the entire southern sky \citep{Wolf2018}. The first all-sky data release of \smp\ (DR1) covers $20,200\,\rm{deg}^2$ of the sky, with almost 300 million detected stellar and non-stellar sources.

The CMD of stars in the Magellanic Clouds region 
within $10^{\circ}$ of $(R.A., Dec.) = (81^{\circ}.91,-69^{\circ}.87)$, the LMC centre from \citet{VanderMarel2002}, is obtained from the slightly updated \smp\ DR1.1 with the following photometric quality selections: 
\begin{gather}
nimaflags = 0, \notag \\  
flags = 0, \notag \\
ngood > 1, \notag \\
ngood\_min > 1\ and\notag \\
nch\_max = 1
\end{gather}
Fig.~\ref{Fig:smp_cmd} shows the resultant CMD, noting that  we have excluded some foreground stars based on their {\it Gaia} parallax (see below).\footnote{The \smp\ photometric data have been de-reddened using the \citet{Schlegel1998} extinction map with the correction by \citet{Schlafly2011}.} 
A number of features appear in the CMD, with hot young stars dominating at bluer colours, whilst RGB stars dominate in the red. In addition, there is a prominent sequence of extremely red stars with $(g-r)_0 > 1.2\ \magt$, and $g_0 \approx 17\ \magt$; this  we identify as Carbon Star candidates in the LMC. Fig.~\ref{Fig:carbon_sky} presents the total sample of stars with $(g-r)_0 > 1.2\ \magt$ from \smp; this map reveals that the main concentrations other than in the Galactic plane are Carbon Stars in the Large and Small Magellanic Clouds (orange points).

\begin{figure}
    \includegraphics[width=\columnwidth]{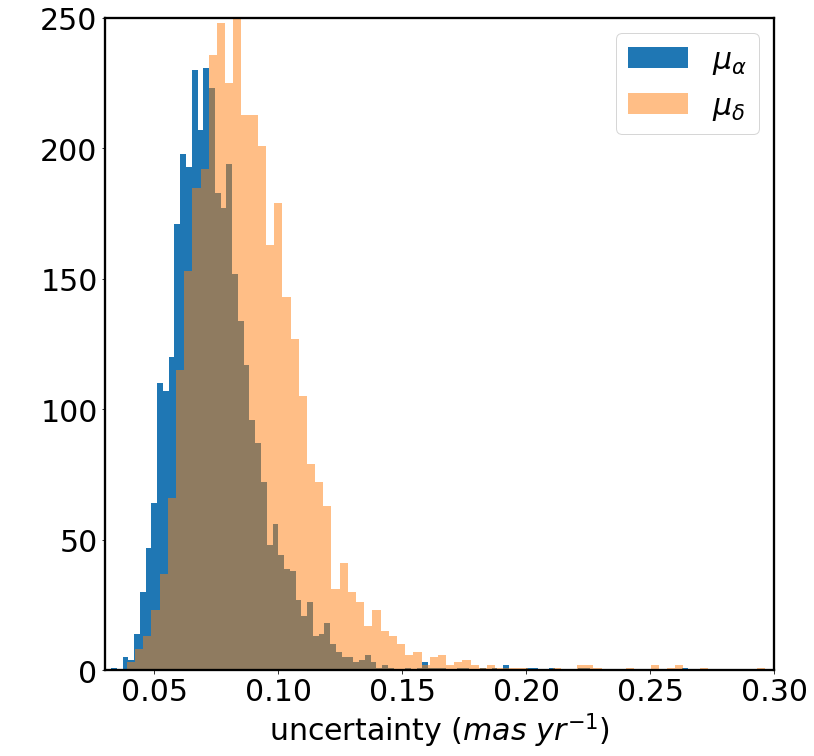}
        \caption{The uncertainty in {\it Gaia} DR2 proper motion for the selected Carbon Star candidates, with a typical uncertainty of $\sim0.07\masyr$. }
        \label{Fig:pm_uncer}
\end{figure}

We confirm the Magellanic Carbon Star candidates through two approaches. Firstly, we take stellar spectra from `The X-Shooter Spectral Library' \citep{Chen2014} and integrate over the \smp\ filter transmission curves \citep{Bessell2011}, to obtain the expected \smp\ colour for different stellar types. In Fig.~\ref{Fig:gr_17}, we find Carbon Stars from this spectral library, marked as ``C'', are closely aligned with the candidate LMC and SMC Carbon Stars from \smp. Secondly, we cross-matched all \smp\ stars in the LMC region with the LMC Carbon Star catalogue of \citet{Kontizas2001}. Fig.~\ref{Fig:matched_carbon} presents the CMD of the matched Carbon Stars. Comparing their catalogue to our selected Carbon Star sample, we find excellent consistency between the two groups. Most Carbon Stars have \smp\ colour $(g-r)_0 > 1\ \magt$ and if we assume that the distance modulus of the LMC is $18.5\ \magt$, the typical absolute magnitude is $M_g \approx -1\ \magt$. 
    
We selected 3578 candidate LMC Carbon Stars from \smp\ in total. These stars extend up to $\approx 9\ \kpc$ from the LMC centre (see Sec.~\ref{sec:result} and Fig.~\ref{Fig:profile}). Although these are rare objects, their high luminosity, and the fact that a simple colour-cut essentially remove all Galactic contamination, together means that they constitute an excellent sample for tracing the dynamical properties of the LMC.

Since the LMC is a highly complex galaxy with a very extended star-formation history, we also consider  RGB stars (selected using the green region in Fig. \ref{Fig:smp_cmd}) and upper main sequence stars (selected from the magenta region in Fig. \ref{Fig:smp_cmd}) as complementary tracers of ancient and young stellar populations, respectively. While the Carbon Stars constitute the primary data set for our analysis, these additional samples allow us to explore variations in the dynamical properties of stellar populations in the LMC.

We cross-match each of the three samples (Carbon Stars, RGB stars, and upper MS stars) with {\it Gaia} DR2 \citep{Gaia1,Gaia2} to obtain astrometric information. The data quality of the current release (DR2) is insufficient to detect parallax precisely at the distance of the LMC, and hence we removed any sources with parallaxes inconsistent with zero at $3\sigma$ as foreground contaminants (cf. Fig.~\ref{Fig:smp_cmd}).

In Fig.~\ref{Fig:pm_uncer} we present the distribution of measurement uncertainties in the two proper motion components for our Carbon Star sample. The typical uncertainty for these stars in {\it Gaia} DR2 is $\sim0.07\masyr$. At the distance of the LMC, this uncertainty roughly corresponds to $\sim 16.5\ \kms$.

\begin{figure*}
    \includegraphics[width=\textwidth]{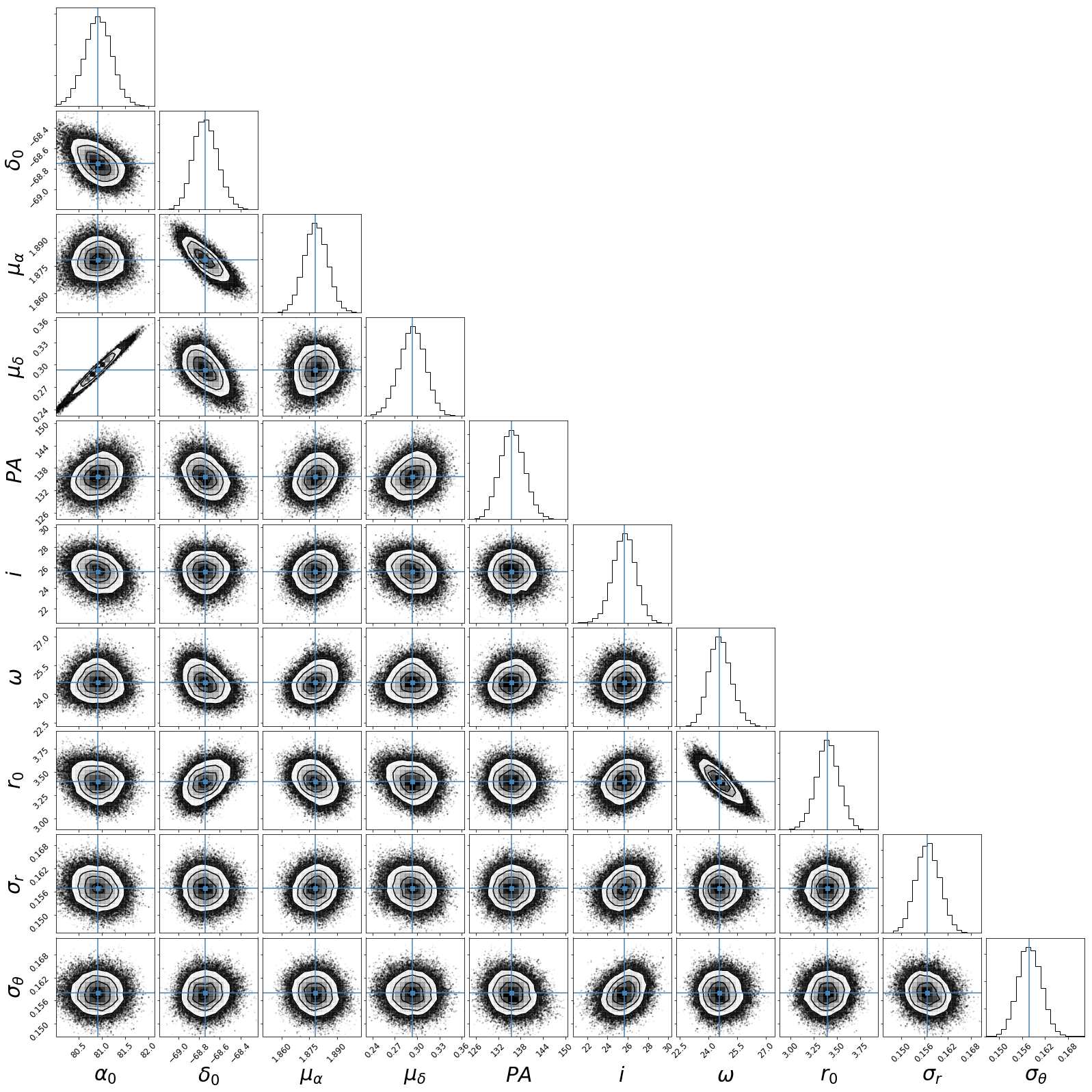}
        \caption{Corner plot summary of the MCMC sampling result for the Carbon Stars. 
        A significant correlation is evident between the inferred proper motion and the inferred dynamical centre of the LMC.
        Additionally, the parameters $\omega$ and $r_0$ are correlated, as $\omega 
        \times r_0$ represents the flat rotation velocity. This figure (as well as Fig. \ref{Fig:mcmc_configure_fit_P4} and Fig. \ref{Fig:mcmc_configure_fit_P1}) is made with the {\it corner} package \citep{cornerplot}}
        \label{Fig:mcmc_configure_fit}
\end{figure*}

\begin{figure*}
    \includegraphics[width=\textwidth]{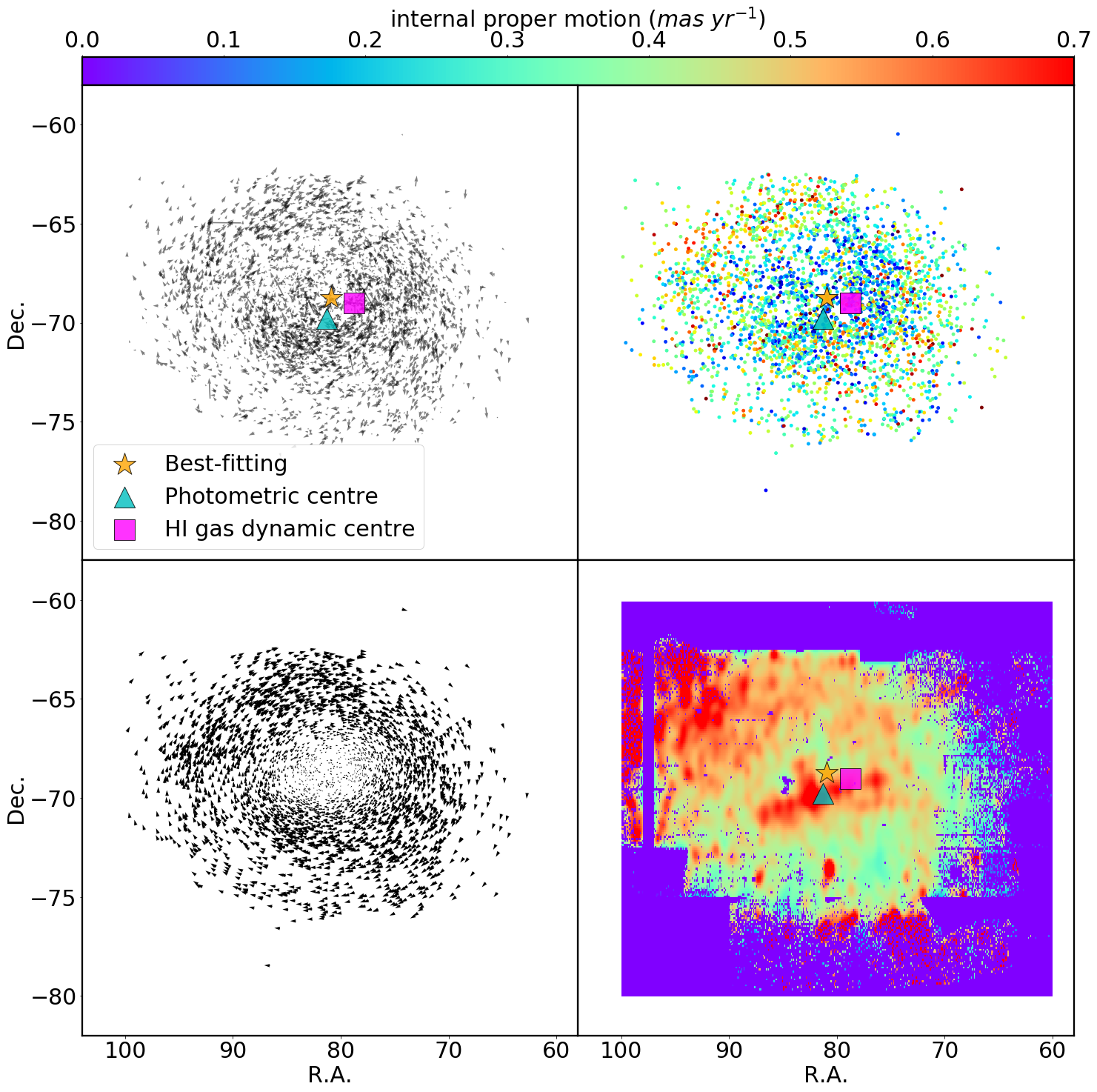}
    
        \caption{{\it Left}: Top: The proper motion of Carbon Star candidates with the bulk proper motion of the LMC subtracted, clearly demonstrating rotation around the LMC centre. 
        We mark our best-fitting centre, the photometric centre and the H{\small\uppercase\expandafter{\romannumeral1}} dynamical centre with the symbols noted. 
        Clearly the best-fitting stellar dynamical centre is offset from both the photometric centre \citep{VanderMarel2001} and the H{\small\uppercase\expandafter{\romannumeral1}} dynamical centre \citep{Luks1992}. Bottom: The proper motion from our model. The model qualitatively matches the observed rotation, though the dispersion also present in the observational data is not reproduced in this panel. 
        {\it Right}: The distribution of Carbon Stars (top), colour coded with their proper motion, and the proper motion heat map of RGB stars (bottom) with pixel size equal to $1^{\circ}\times1^{\circ}$, which has been smoothed with a Gaussian kernel of 3 pixels in size. In the RGB star heat map, the stellar bar clearly stands out with high proper motion at the LMC photometric centre. However, the Carbon Stars show no significant similar feature. This comparison indicates that the Carbon Stars are more likely to be located in the disk rather than the bar of the LMC. }
    \label{Fig:propermotion}
\end{figure*}

\section{Results}
\label{sec:result}

\subsection{Kinematics}
The observed proper motions of the stars in LMC consist of the bulk motion and their internal motion relative to the LMC system.
Because the LMC has a large angular size, variations in viewing perspective mean that the apparent contribution due to the bulk motion changes as a function of position on the sky. Therefore, we model the motions of the stellar sample as:
\begin{equation}
	\textbf{V} = \textbf{V}_{bulk} + \textbf{V}_{inter}
\end{equation} 
Here $\textbf{V}$ is the 3-D velocity, while $\textbf{V}_{inter}$ describes the internal velocities and $\textbf{V}_{bulk}$ is the bulk motion of the LMC. The latter has two components in proper motion that we set as free parameters, and one line-of-sight component that we fix as $262.2\ \kms$\citep{VanderMarel2002}.
In modelling the internal velocity components, we assume a simplified model where the Carbon Stars are in a thin rotating disk where the $x$, $y$,$z$ coordinate system centred at the LMC centre and the disk lies on the $xy$ plane. The rotation curve is given by:
\begin{gather}
	V_{\phi} = \omega\ r ,\ {\rm for}\ r < r_0 \notag \\
	V_{\phi} = \omega\ r_0 ,\ {\rm for}\ r \ge r_0
\end{gather}
where $r = \sqrt{x^2 + y^2}$ is the de-projected in-plane radius; $r_0$ is the break radius; $\omega$ is the constant angular speed of the inner regions of the LMC; $\phi$ is the in-plane directional angle and $V_{\phi}$ is the rotational speed. 
In this simple model, we assume no net velocity components in the $z$ and $r$ directions, and consider a constant in-plane radial ($\sigma_{r}$) and a tangential velocity dispersion ($\sigma_{\phi}$).

We then project the total velocity into the sky by assuming that the LMC disk is in a plane tilted with respect to the line of sight. The dynamical centre, denoted as $\alpha_0$ and $\delta_0$, are set as free parameters, with the distance to the LMC centre assumed to be $49.9\kpc$ \citep{deGrijs2014}. To define the 3D velocity, we need two additional quantities: the position angle $\theta$ \cite[from North to East as suggested in][]{VanderMarel2002} of the line of nodes and the inclination angle $i$. These are free parameters in our model. We define our transformations as:
\begin{flalign}
	\textbf{V}_{sky}& =  \mathbb{R}_{x}(i)\cdot\mathbb{R}_{z}(\theta - \pi) \cdot\mathbb{R}_{x}(\delta_{0} - \frac{\pi}{2})\cdot\mathbb{R}_{z}(\alpha_{0} - \frac{\pi}{2})\cdot\textbf{V} \notag \\
	\mu_{\alpha}& = -V_{sky}(x)\sin(\alpha) + V_{sky}(y)\cos(\alpha)  \notag \\
	\mu_{\delta}& = V_{sky}(z)\cos(\delta)  -(V_{sky}(x)\cos(\alpha) + V_{sky}(y)\sin(\alpha))\sin(\delta) 
\end{flalign}
where $\mathbb{R}$ is the rotation matrix along the corresponding axis, $\alpha$ and $\delta$ is the sky position of each Carbon Star in the sample and $V_{sky}(x)$ is the component of $\textbf{V}_{sky}$ in Cartesian coordinates. This model builds up a correlation between in-plane velocity and proper motion of stars as a function of star position\footnote{See Appendix \ref{app:model} for details of the correlation and the corresponding Jacobian matrix and determinant}. We explore the likelihood space with an MCMC sampling algorithm \citep{Foreman-Mackey2013} to find the best-fitting model and corresponding uncertainties on parameters. Fig.~\ref{Fig:mcmc_configure_fit} shows the corner plot summary of the MCMC sampling results. The best parameter values are:
\begin{flalign}
    \alpha_0 &= 80.90^{\circ}\pm0.29,\ &\delta_0 &= -68.74^{\circ}\pm0.12 \notag \\
    \mu_{\alpha} &= 1.878\pm0.007\ \masyr,\ &\mu_{\delta} &= 0.293\pm0.018\ \masyr \notag \\
    \theta &= 135.6^{\circ}\pm3.3,\ &i &= 25.6^{\circ}\pm1.1  \notag \\
    \omega &= 24.6\pm 0.6\ \kms\ \kpc^{-1},\ &r_0 &= 3.39\pm 0.12\ \kpc\notag \\ 
    \sigma_r &= 0.157\pm 0.003\ \masyr,\ &\sigma_{\phi} &= 0.158\pm 0.003\ \masyr
\end{flalign}
We summarise the LMC parameters from our best-fitting configuration model and compare them to literature values in Tab.~\ref{Tab:parameter}. Here we also list the fitting results from the RGB stars and MS stars; see the detailed discussion on Sec.\ref{sec:multi-pop}

\begin{table*}

	\begin{tabular}{*{5}{c}}
		\hline
		Dynamical Centre & Bulk Motion $(\mu_{\alpha*}, \mu_{\delta})$ / $mas\ yr{-1}$ & $\theta\ /\ ^{\circ}$ & $i\ /\ ^{\circ}$ & reference \\
		\hline
		$(80.90\pm0.29, -68.74\pm0.12)^{a}$ & $(1.878\pm0.007,0.293\pm0.018)^{a}$ & $(135.6\pm3.3)^{a}$ & $(25.6\pm1.1)^{a}$ & Carbon Stars in this work\\
		$(81.23\pm0.04, -69.00\pm0.02)$ & $(1.824\pm0.001,0.355\pm0.002)$ & $(134.1\pm0.4)$ & $(26.1\pm0.1)$ & RGB Stars in this work\\
		$(80.98\pm0.07, -69.69\pm0.02)$ & $(1.860\pm0.002,0.359\pm0.004)$ & $(152.0\pm1.0)$ & $(29.4\pm0.4)$ & Young MS in this work\\
		$(81.91 \pm 0.98,-69.87 \pm 0.41)$ & - & $129.9 \pm 6.0$ & ${34.7 \pm 6.2}^b$ & \citet{VanderMarel2002} \\
		$(81.91 \pm 0.98,-69.87 \pm 0.41)^c$ & - & $142 \pm 5$ & ${34.7 \pm 6.2}^b$ & \citet{Olsen2011}\\
		$(78.76 \pm 0.52,-69.19 \pm 0.25)$ & $(1.910 \pm 0.020,0.229 \pm 0.047)$ & $147.4 \pm 10.0$ & $39.6 \pm 4.5$ & \citet{VanderMarel2014}\\
		$(80.78, -69.30)^{e} $ & - & $150.76 \pm 0.07$ & $25.05 \pm 0.55$ & \citet{Inno2016}\\
		$(78.77, -69.01) ^d$ & $(1.850 \pm 0.030,0.234 \pm 0.030)$ & $[106.4,134.6]^{f}$ & $[30.1,61.5]$ & \citet{Helmi2018} \\
		$(82.25, -69.5)^{b}$ & - & $149.23^{+6.43}_{-8.35}$ & $25.86^{+0.73}_{-1.39}$ & \citet{Choi2018} \\
		$(81, -69.75)$ & - & $[130,135]$ & $[32,35]$ & \citet{Vasiliev2018} \\
		\hline
	\end{tabular}
		\caption{The LMC reference or fitted parameters from our best-fitting model and from the literature. The first column is the centre point; the second column is the corresponding bulk motion of the LMC; the third column is the position angle of the line of nodes; the fourth column is the inclination angle.
			$^a$ The uncertainties are from MCMC sampling,  
			$^b$ Taken from \citet{VanderMarel2001}, 
			$^c$ Taken from \citet{VanderMarel2002}, 
			$^d$ Taken from H{\small\uppercase\expandafter{\romannumeral1}} centre \citep{Luks1992}, 
			$^e$ Cepheids centroid in \citet{Inno2016},
			$^f$ See \citet{Helmi2018} for a detailed model discussion.
			}
	\label{Tab:parameter}

\end{table*}

The bulk motions from the differing measurements are in general agreement.We note that the inferred bulk proper motions are correlated with the assumed dynamical centre (cf. Figure \ref{Fig:mcmc_configure_fit}). This means that an identical intrinsic true-space motion will result in a varied measurement of the proper motion, if the reference centre under consideration is different. This is the main cause of the difference between each results. For example, as \citet{Helmi2018} suggests, if the dynamical centre were fixed as the photometric centre from \citet{VanderMarel2001}, then the their proper motion would be $(1.890,0.314)\mathrm{mas\ yr^{-1}}$. For Carbon Stars, the best-fitting configuration parameter $\theta$ agrees well with other works listed in the table, and $i$ agrees closely with the purely geometric measurement from \citet{Inno2016} and \citet{Choi2018} . 

In the following analysis, unless specified otherwise, the bulk motion has been subtracted from all velocities using our best-fitting proper motion results and the assumed line-of-sight velocity. In the left panel of Fig.~\ref{Fig:propermotion}, we see that the residual proper motions for the Carbon Stars indicate they are clearly rotating around the LMC centre. Also shown in this panel, our inferred dynamical centre for the Carbon Star sample differs from the H{\small\uppercase\expandafter{\romannumeral1}} dynamical centre, and the photometric centre of the LMC.

\begin{figure}

  \centering
  \includegraphics[width=\columnwidth]{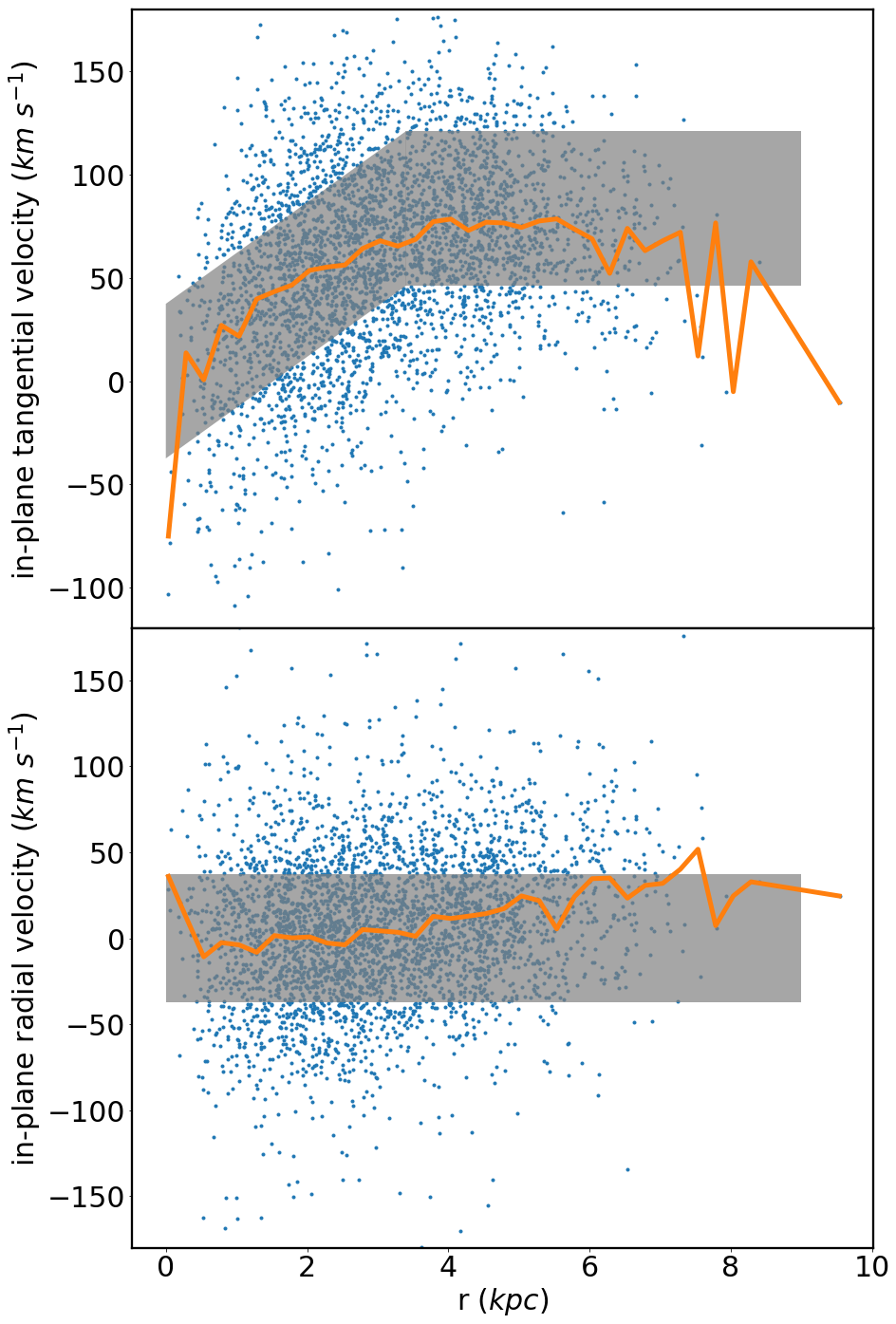}
\caption{The rotational (top) and radial (bottom) speed profiles as a function of radius for Carbon Stars and the best-fitting kinematic model, indicating an asymptotic flat rotation speed of $V_{rot} = 83.6\pm 1.7 \kms$. The radial speed is scattered around $0$, with the grey shading indicating the best-fitting dispersion as a function of radius. The orange lines indicate the average velocity within $0.25\ \kpc$ bins. }
\label{Fig:profile} 
\end{figure}

In the right panel of Fig.~\ref{Fig:propermotion} we show the distribution of proper motions for our Carbon Stars compared to those for the RGB sample (selected from the \smp\ CMD as shown in Fig.~\ref{Fig:smp_cmd}). The bar clearly stands out in the map of RGB stars with higher proper motion, whereas this is not clear in the Carbon Star distribution, suggesting that the Carbon Stars are generally not drawn from the bar. Interestingly, \citet{Olsen2011} suggests that part of the Carbon-rich AGB stars are likely to be counter-rotating or form a second disk that have different inclination. They also found those stars have a distribution that avoids the LMC bar.

Fig.~\ref{Fig:profile} shows the tangential ({\it{top}}) and radial ({\it{bottom}}) velocity profile. In both panels, the gray shaded region represents the best fitting model, and the widths of the gray region represent the dispersions. As inferred in the top panel of this figure, the rotation speed gradually increases inside $\sim 3\ \kpc$, in agreement with the results of \citet{Helmi2018}. After this point, the rotation speed flattens to $V_{rot} = 83.6\pm 1.7 \kms$, corresponding to a proper motion of $\sim 0.353\masyr$. The average tangential speed may exhibit a mild decrease at large radii (beyond $\sim 7 \kpc$), although this conclusion is only tentative due to the paucity of data in this region. The bottom panel in Fig.~\ref{Fig:profile} shows that the radial speed averages to approximately zero, with a slightly increasing tendency outwards.

\subsection{Mass}
\label{sec:mass}

The total LMC mass is known to be a key factor for the first-infall scenario, where a massive LMC ($> 1\times10^{11}\ \msun$) is required to ensure that the Clouds evolved as a bound pair for at least 5 Gyr to form the Magellanic Stream \citep{Besla2012}. As shown in \citet{Kallivayalil2013}, the first-infall scenario becomes more likely for massive LMC; a massive LMC moreover implies a relatively rapid merger with the Milky Way $\sim 2.5$\ Gyr from now \citep{Cautun2018}.

Whilst the LMC is thought to possess a massive dark matter halo, a lack of dynamical tracers at very large radii has limited the determination of the total LMC mass through kinematic means. Instead, the total mass is generally inferred through more indirect methods. For instances, \citet{Penarrubia2016}, considering the timing argument within the Local Group, proposed an infall LMC mass equal to $(2.5\pm0.9)\times10^{11}\ \msun$; \citet{Cautun2018} showed that in the EAGLE cosmological simulations, LMC-mass satellites with an SMC-like companion typically have a total halo mass of $(3.0^{+0.7}_{-0.8})\times10^{11}\ \msun$. \citet{Erkal2018} inferred the total LMC mass to be $(1.38^{+0.27}_{-0.24})\times10^{11}\ \msun$ from the observed perturbation on the Orphan stellar stream in the Milky Way halo; \citet{Erkal2019} estimated a lower limit of $1.24\times10^{11}\ \msun$ in order to bind the six most-likely infalling dwarf companions of the Magellanic system, and \citet{Belokurov2019} showed that simulations with a low LMC mass ($2\times10^{10}\ \msun$) can better explain the observed northern spiral structure of the LMC, partly as a consequence of the recent interaction with the Milky Way. 

From our best-fitting model, we estimate the LMC mass by considering the circular velocity and adopting the approach discussed in \citet{VanderMarel2002}: $V_{circ}^{2} = V_{rot}^{2}+\kappa\sigma_{rad}^{2}$, where $\kappa = 6$ and $\sigma_{rad}$ is the dispersion of the radial velocity profile, which together constitute the {\it asymmetric drift} correction. Noting that our best-fitting values of $V_{rot} = 83.6\pm 1.7 \kms$, and the radial dispersion $\sigma_{rad} = 0.157\pm 0.003\ \masyr$, corresponding to $37.1\ \kms$, we obtain a circular velocity equal to $V_{\rm{circ}} \sim123.6\pm1.9\ \kms$ at $7\ \kpc$. 
Using the equation $M = V_{circ}^{2}r/G$ and $G = 4.3007\times10^{-6}\ \kpc\ (km\ s^{-1})^{2}\ \msun^{-1}$,  we estimate an upper limit on LMC mass within $7\ \kpc$ to be $M_{\rm LMC}(<7\kpc) = (2.5\pm0.1)\times10^{10}\msun$. This mass agrees with other estimations based on stellar dynamics, e.g., \citet{VanderMarel2014, VanderMarel2002}. \citet{VanderMarel2014} estimated the tidal radius to be $22.3 \pm 5.2\ \kpc$, whilst \citet{Navarrete2019} found stars that match the expected velocity gradient for the LMC halo extending up to $29\ \kpc$ away from the LMC centre. If we assume that the circular velocity remains constant out to $30 \kpc$, the mass within tidal radius is $(1.06 \pm 0.32)\times10^{11}\ \msun$. 

Whilst we note the lack of data at large radii,  both the velocity dispersion and the tangential velocity tentatively exhibit a decreasing tendency with radius in Figure \ref{Fig:profile} \citep[see also the dispersion profile in ][]{Vasiliev2018}. If true, this would imply that our total extrapolated mass within an assumed tidal radius of $30\ \kpc$ is likely to be an upper limit. However, determining the tidal radius is difficult, 
 so the total mass would be an approximate estimation. In the future---potentially during {\it Gaia} DR3 era---the tidal radius can be more accurately determined, yielding a better dynamical mass estimation. Generally, this result matches  recent mass estimations from e.g. \citet{VanderMarel2014,Erkal2018,Erkal2019}, while smaller than (or at the lower end of) the LMC mass from e.g. \citet[][]{Penarrubia2016,Cautun2018,Shao2018,Garavito-Camargo2019}. As previously noted, the mass of the LMC is crucial for understanding its evolutionary history. Not only does the first in-fall scenario require the LMC to be larger than $\sim 1\times10^{11}\ \msun$ \citep{Kallivayalil2013}, a lower mass LMC also leads to a much later LMC-MW merger \citep{Cautun2018} and produces substantially less perturbation in the Milky Way halo during its in-fall \citep{Garavito-Camargo2019}.

\begin{figure}
    \includegraphics[width=\columnwidth]{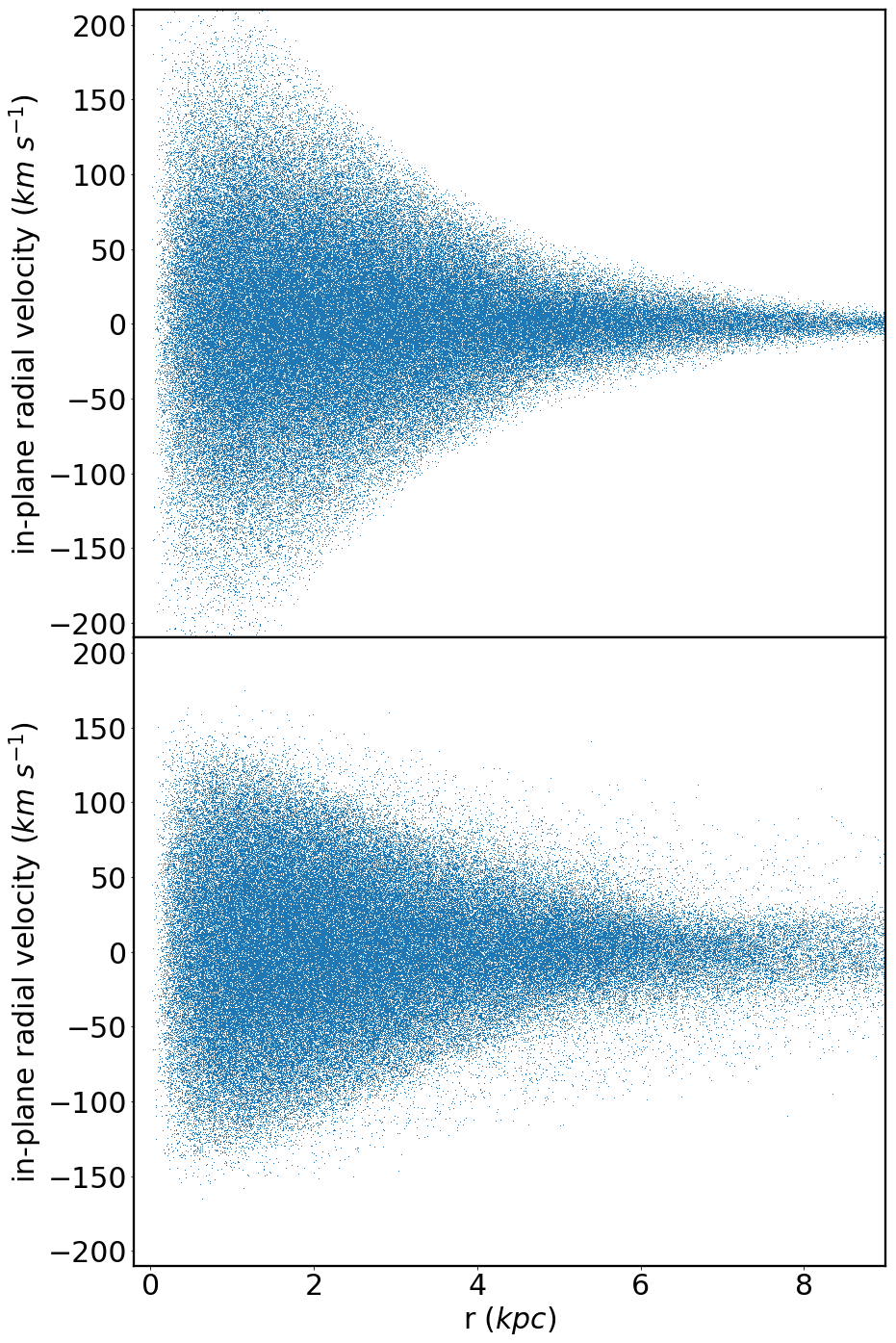}
        \caption{The radial profile of the initial/final (top/bottom panel) state of LMC-SMC interaction simulation \citep[see][]{Guglielmo2014}. The LMC-SMC interaction increases the dispersion and this effect is obvious at the out-skirt of the LMC.}
        \label{Fig:radius_rotation_simulation}
\end{figure}

\subsection{Velocity Dispersions}
As noted above, and described in detail in the Appendix, our model for the velocity properties of the Carbon Star sample in the LMC also incorporates the velocity dispersion in the rotational and radial directions, found to be $(\sigma_r,\sigma_{\theta}) = (0.157\pm 0.003,\ 0.158\pm 0.003)\ \masyr$ corresponding to $\sim 37 \kms$ (Fig.~\ref{Fig:profile}), which is comparable with the dispersion in the inner LMC derived from \citet{Vasiliev2018}. This dispersion significantly contributes to the total mass estimate via the {\it asymmetric-drift} correction \citep[e.g.][ and our estimation in Sec.~\ref{sec:mass}]{VanderMarel2002,Dehnen1998}. 

It is possible that dynamical interactions between the LMC and SMC, the most recent of which likely occurred $\sim100-200\ \rm{Myr}$ ago, have had a substantial effect on the velocity dispersion in the LMC. 
Interactions between the LMC and SMC are supported by several lines of evidence. For example, in the SMC, the gas outflow found by \citet{McClure-Griffiths2018} and the shell of young stars recently studied by \citet{Martinez-Delgado2019} are both indicative of possible interactions with the LMC. For the LMC, \citet{Choi2018} recently identified an outer warp in the disk, and a tilted bar, using red clump stars, consistent with a close encounter with the SMC \citep[see also,][etc]{Besla2012, Noel2013, Guglielmo2014,Carrera2017,Zivick2018}. \citet{Joshi2019} found a common enhancement of the Cepheid population in both the LMC and SMC, suggesting an interaction $\sim 200\ \rm{Myr}$ ago between the Clouds. \citet{Schmidt2018} found the stars in the Magellanic Bridge are moving towards the LMC, supporting the idea that they, or the gas from which they formed, has been stripped from the SMC due to dynamical interactions. Finally, \citet{Olsen2011} suggested that a proportion of the carbon-rich AGB stars in the LMC may have come from the SMC. If this is correct then it is possible that there may be a non-disk, e.g., stripped SMC, component in our sample, potentially explaining in some part the observed spatial and kinematic offsets identified previously and inflating the observed dispersion.

We illustrate the effect of LMC-SMC interaction on the dispersion profile withe the $3\ \Gyr$ snapshot model from \citet{Guglielmo2014}. The LMC and SMC have two close encounters during the integration. This simulation adopts an LMC mass of $1.9\times10^{10}\ \msun$ within $9\ \kpc$, which is roughly comparable with, but somewhat smaller than, our result. Despite this mild discrepancy (see Sec.~\ref{sec:mass}), the simulation should provide an indicative picture of the effect of LMC-SMC interactions. 

Fig.~\ref{Fig:radius_rotation_simulation} shows the initial and final state of the radial velocity profiles. The interaction between the LMC and SMC clearly increases the dispersion, which is more apparent in the out-skirts of LMC. For example, at $5\ \kpc$ the initial state has a tangential velocity dispersion of $\sim 23.7\ \kms$, which increases to $\sim 33.3\ \kms$ at the final state. Simulations of an isolated LMC do not reproduce the observed dispersion, suggesting that this is not due to the natural evolution of the LMC. However, the LMC-SMC interaction model cannot fully reproduce the radial velocity profile (cf Figure.\ref{Fig:profile}), suggesting the current profile cannot be simply explained by LMC-SMC interactions alone. An alternative explanation is presented in \citet{Armstrong2018}, based on the results in \citet{Olsen2011}, the authors discussed the possibility that a third dwarf galaxy merging with LMC might have caused an increase in the velocity dispersion. However, this remains an open question.

\subsection{Multi-population analysis}
\label{sec:multi-pop}
So far we have presented an analysis based on Carbon Stars from which we estimated the dynamical properties and the mass of the LMC. As previously described, we also identified two additional sets of stars: upper MS stars, and RGB stars (see Fig.~\ref{Fig:smp_cmd}). The dynamical properties of these stars could potentially be different from those of the Carbon Stars since they trace populations of different ages and, therefore, have likely experienced different evolutionary histories. For example, we see that the bar clearly stands out with higher proper motion in the RGB sample but not in the Carbon Star sample. Furthermore, the upper MS stars formed only relatively recently and may therefore still retain a signature of their formation conditions rather than being fully mixed with older populations. 

To characterise the dynamical properties of each population, we apply the same algorithm to the young MS stars and RGB stars as we do for the Carbon Star population. However, since there are more contaminants in the RGB and MS samples compared to the Carbon Stars, we exclude outliers by adding selection constraints on the proper motions: $1 < PM_{R.A.} < 2.5\ \masyr$ and $-1 < PM_{Dec.} < 1.5\ \masyr$. Tab.~\ref{Tab:multi-pop} summarises the best-fitting parameters for each population, with Fig.~\ref{Fig:mcmc_configure_fit_P4} and Fig.~\ref{Fig:mcmc_configure_fit_P1} showing the corresponding parameter distributions for the RGB and MS stars. Our results suggest that the inferred bulk proper motions and the estimated circular velocities for the three populations roughly agree with each other.

The inferred inclination angles are relatively similar for all three populations; however the $PA$ for the young MS stars is significantly different to that for the RGB and Carbon stars. Interestingly, this inclination is in good agreement with the inclination estimation from Red Clump ($141.5\pm4.5$ from \citet{Subramanian2013} and $149.23\pm8.35$ from \citet{Choi2018}), RR Lyrae ($150.76\pm0.07$ from \citet{Inno2016}), and especially young stars ($147.4\pm10$ from \citet{VanderMarel2014}). In addition, the rotation profile parameters $\omega$ and $r_0$ for Carbon Stars and RGB stars agree quite closely, but are rather different to those for the young MS stars, indicating the dynamics of young MS stars in the central regions of the LMC are indeed different to those for older stars.

\begin{figure}
    \includegraphics[width=\columnwidth]{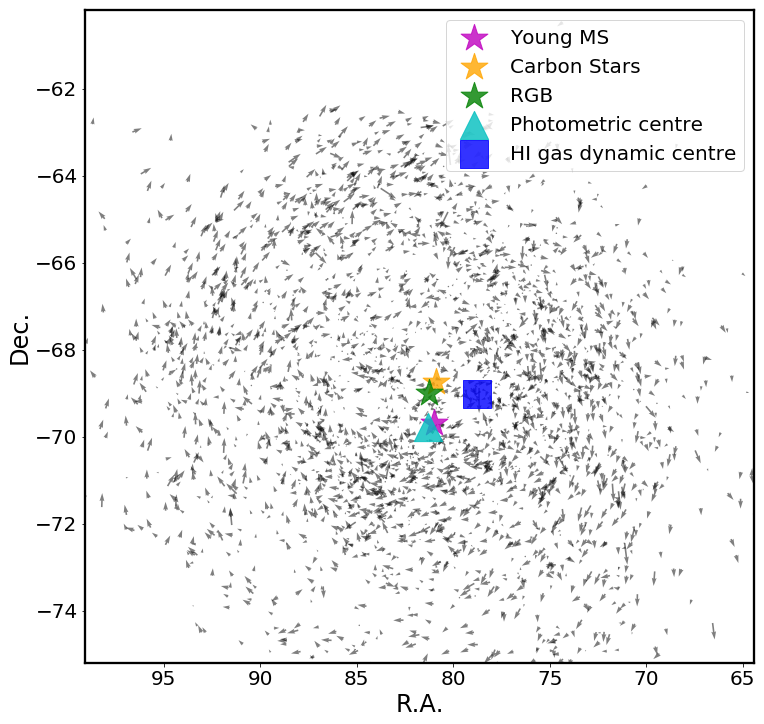}
        \caption{The best-fitting centre for young MS stars, Carbon Stars and RGB stars, compared to the photometric centre \citep{VanderMarel2001} and photometric centre \citep{Luks1992}. On the background is the internal proper motion map of Carbon Stars. The dynamical centre of the young MS stars is close to the photometric centre. The dynamical centre of the RGB stars and Carbon Stars are close to each other, but roughly $1^{\circ}$ away from the young MS star centre and the photometric centre.}
        \label{Fig:multi-centre}
\end{figure}

The most striking difference between the young and old populations is in the inferred dynamical centres. The best-fitting dynamical centre for the young MS stars is $\sim 1^{\circ}$ away from the dynamical centre for the Carbon Stars and $0.7^{\circ}$ away from the dynamical centre for the RGB stars, while the centres for the latter two population are very close to each other. In Fig.~\ref{Fig:multi-centre} we show the best-fitting dynamical centres for the three populations, compared to the photometric centre \citep{VanderMarel2001} and the H{\small\uppercase\expandafter{\romannumeral1}} dynamical centre \citep{Luks1992}. The centre for the young MS stars is close to the photometric centre ($0.1^{\circ}$), which is at the centre of the bar, and $1.19^{\circ}$ away from the H{\small\uppercase\expandafter{\romannumeral1}} dynamical centre. Given the sample size of 3000 stars, tests using mock data indicate that our measurements are robust to within $\approx 0.2^{\circ}$, which agrees with the fitting results from Carbon Stars; moreover, since we apply the same algorithm to each stellar population, model-dependence cannot be the cause of the observed offsets.

The difference between the dynamical centres for the young stars and the H{\small\uppercase\expandafter{\romannumeral1}} gas out of which they presumably formed, is intriguing. One possibility is that the most recent LMC-SMC interaction, if it occurred after the majority of the young stars had formed (i.e., within the last $\sim100-200$\ Myr) could have substantially perturbed the H{\small\uppercase\expandafter{\romannumeral1}} relative to the stars. It is also plausible that if the formation of the young stars was in fact triggered by an LMC-SMC interaction, that this star formation may not have been uniform within the H{\small\uppercase\expandafter{\romannumeral1}}, leading to an apparent discrepancy in their dynamical centres. A final possibility relates to additional forces felt by the gas compared to the stars, as a consequence of ram pressure due to the Milky Way's hot corona. For example, \citet{Belokurov2017} showed that RR Lyrae stars and the H{\small\uppercase\expandafter{\romannumeral1}} gas in the Magellanic Bridge -- although both ostensibly stripped from the SMC -- possess quite different spatial distributions, an observation they attribute to the effects of the Milky Way's corona. That the inferred bulk proper motion of the LMC is largely towards the east, whereas the dynamical centre of the H{\small\uppercase\expandafter{\romannumeral1}} gas sits to the west of that for the young stars, is consistent with this interpretation.

\begin{table*}

	\begin{tabular}{*{7}{c}}
		\hline
		Population & $\alpha_0\ /\ ^{\circ}$ & $\delta_0\ /\ ^{\circ}$ & $\mu_{\alpha}\ /\ \masyr$ & $\mu_{\delta}\ /\ \masyr$ & $PA\ /\ ^{\circ}$ & $i\ /\ ^{\circ}$\\
		\hline
		Carbon Stars & $80.90 \pm 0.29$ & $-68.74 \pm 0.12$ & $1.878 \pm 0.007$ & $0.293\pm0.018$ & $135.6\pm3.3$ & $25.6\pm 1.1$\\ 
		RGB Stars & $81.23 \pm 0.02$ & $-69.00 \pm 0.01$ & $1.824 \pm 0.001$ & $0.355\pm0.002$ & $134.1\pm0.4$ & $26.1\pm 0.14$ \\ 
		Young MS stars & $80.98 \pm 0.08$ & $-69.69 \pm 0.02$ & $1.860 \pm 0.002$ & $0.359\pm0.005$ & $152.0\pm1.0$ & $29.4\pm 0.45$ \\ 
		\hline
		Population & $\omega\ /\ \kms\kpc^{-1}$ & $r_0\ /\ \kpc$ & $\sigma_r\ /\ \masyr$ & $\sigma_{\theta}\ /\ \masyr$  &  $V_{Circ}\ /\ \kms$ & \\
		\hline
		Carbon Stars &  $24.6 \pm 0.6$ & $3.39\pm 0.12 $ & $0.157\pm 0.003 $ & $0.158\pm 0.003 $ & $123.6 \pm 1.9$ & \\
		RGB Stars & $23.3 \pm 0.1 $ & $3.14\pm 0.02 $ & $0.183\pm 0.001 $ & $0.170\pm 0.001 $ & $128.9 \pm 0.3$ & \\
		Young MS stars & $38.5 \pm 0.6 $ & $1.84\pm 0.03 $ & $0.174\pm 0.001 $ & $0.156\pm 0.001$ & $122.9 \pm 0.7$ & \\
		\hline
	\end{tabular}
		\caption{This table summaries the best-fitting parameters for the Carbon Stars, RGB stars and Young MS stars. Note that here we assume the same $\kappa$ for the {\it asymmetric drift} correction when calculating the circular velocity (see Sec.\ref{sec:mass}).}
	\label{Tab:multi-pop}

\end{table*}

\section{Conclusions}
\label{sec:conclusion}
In this work, we select 3578 Carbon Stars candidates from \smp DR1.1, including parallax information from {\it Gaia} DR2 to provide additional robustness; these Carbon Stars have very red $g-r$ colours, which are easily isolated using the $g-r$ vs $g$ CMD. In addition, we also consider young MS and RGB stellar samples. From a comparison with a map of RGB stars, we note that the Carbon Star candidates are more likely located in the LMC disk, not showing the prominent bar features seen in the RGBs.

By assuming the stars are located and move in the disk, we construct a rotating planar model of the LMC and find the best fitting geometric and kinematic parameters for the Carbon Star sample. The inferred properties of the LMC are in reasonable agreement with previous measurements \citep[e.g.][]{VanderMarel2002,Olsen2011,Kallivayalil2013}. In addition, we find a significant offset between the centre of the Carbon Star sample and both the H{\small\uppercase\expandafter{\romannumeral1}} dynamical centre and the photometric centre of the LMC, a signature that could result from the on-going LMC-SMC interaction. 

We applied the same fitting algorithm to the RGB stars and young MS stellar samples. The PA for the young stars is significantly different to the results from old stars, suggesting that they are drawn from different distributions. The dynamical centre for the RGB stars is close to the Carbon Star centre, and hence exhibits the same offset from the photometric centre. However, the dynamical centre for the young MS stars is close to the photometric centre and is significantly offset from the old populations, indicating that the young stars have different dynamical properties. We speculate the observed offset---between the dynamical centre for the young stars and that for the H{\small\uppercase\expandafter{\romannumeral1}} gas out of which they presumably formed---may reflect the effects of a possible LMC-SMC interaction in the period since the young stars formed, and/or the additional forces felt by the gas compared to the stars, due to ram pressure from the Milky Way's hot corona.

Using a simulation of the LMC-SMC interaction, we illustrate that this can increase the observed velocity dispersion, but further interactions, possibly with a third dwarf galaxy, may be needed to fully account for the observations. Our model contains a constant dispersion and it is weighted by the data. Compared to \citet{Vasiliev2018}, it hence overestimates the dispersion in the velocity profiles at large radii.

From the tangential velocity profile and its dispersion, we measure the circular velocity to be $V_{\rm{circ}} \sim123.6 \pm 1.9\ \kms$ at $7\ \kpc$, implying an LMC mass within $7\ \kpc$ of $(2.5\pm0.1)\times10^{10}\msun$. From this, we estimate the total LMC mass within  $30\ \kpc$ to be $(1.06 \pm 0.32)\times10^{10}\ \msun$ under the assumption of a constant circular velocity to the tidal radius. The radial dispersion significantly contributes to the mass estimation via the {\it asymmetric drift} correction. Since we adopt a model with a constant dispersion, which may consequently overestimate the dispersion at larger radii, the mass we estimate here plausibly represents an upper limit for the LMC mass within $30\ \kpc$, and we note that a better mass estimation would require an accurate tidal radius estimation. The mass determined in this present study, 
whilst significantly smaller than some of very massive LMC models considered in the literature, is consistent with the mass estimation from  tidal-interaction and perturbation considerations \citep[e.g.][]{Erkal2018,Erkal2019}.

\section*{Acnowledgements}
ZW gratefully acknowledges financial support through a the Dean's International Postgraduate Research Scholarship from the Physics School of the University of Sydney. DM holds an Australian Research Council (ARC) Future Fellowship (FT160100206). We thank the anonymous reviewer for their constructive suggestions.

The national facility capability for SkyMapper has been funded through ARC LIEF grant LE130100104 from the Australian Research Council, awarded to the University of Sydney, the Australian National University, Swinburne University of Technology, the University of Queensland, the University of Western Australia, the University of Melbourne, Curtin University of Technology, Monash University and the Australian Astronomical Observatory. SkyMapper is owned and operated by The Australian National University's Research School of Astronomy and Astrophysics. The survey data were processed and provided by the SkyMapper Team at ANU. The SkyMapper node of the All-Sky Virtual Observatory (ASVO) is hosted at the National Computational Infrastructure (NCI). Development and support the SkyMapper node of the ASVO has been funded in part by Astronomy Australia Limited (AAL) and the Australian Government through the Commonwealth's Education Investment Fund (EIF) and National Collaborative Research Infrastructure Strategy (NCRIS), particularly the National eResearch Collaboration Tools and Resources (NeCTAR) and the Australian National Data Service Projects (ANDS).

This work has made use of data from the European Space Agency (ESA) mission
{\it Gaia} (\url{https://www.cosmos.esa.int/gaia}), processed by the {\it Gaia}
Data Processing and Analysis Consortium (DPAC,
\url{https://www.cosmos.esa.int/web/gaia/dpac/consortium}). Funding for the DPAC
has been provided by national institutions, in particular the institutions
participating in the {\it Gaia} Multilateral Agreement.




\bibliographystyle{mnras}
\bibliography{zw_lmc} 



\newpage
\onecolumn
\appendix

\section{LMC model}
\label{app:model}
As noted in Section~\ref{sec:result}, 
we assume the stars in the LMC disk plane are on circular orbits so:
\begin{gather}
	V_{\phi} = \omega\ r,\ {\rm for}\ r < r_0 \notag \\
	V_{\phi} = \omega\ r_0,\ {\rm for}\ r \ge r_0 \notag\\
	\mathbf{V} = (v_x,v_y,v_z) = (-V_{\phi}\sin(\phi),V_{\phi}\cos(\phi),0) + \mathbf{V_{bulk}} \notag 
\end{gather}
Here the $r$ and $\phi$ will depend on the configuration of LMC and are functions of stars' sky position and $V_{bulk}$ is the constant bulk motion. We then assume constant dispersion in both tangential and radial direction:
\begin{equation}
    p(v_{r},v_{\phi}) = \frac{1}{2\pi\sigma_{r}\sigma_{\phi}}\exp{\left[-\left(\frac{(v_{\phi} - V_{\phi}(r))^2}{2\sigma_{\phi}^2} + \frac{{v_r}^2}{2{\sigma_r}^2}\right)\right]} \notag
\end{equation}
Assume the configuration of LMC: position angle $\theta$, inclination $i$ and centre $(\alpha_0, \delta_0)$, we derive the model prediction on the proper motion would be:
\begin{flalign}
	\mathbf{V}_{sky} &=  \mathbb{R}_{x}(i)\cdot\mathbb{R}_{z}(\theta - \pi) \cdot\mathbb{R}_{x}(\delta_{0} - \frac{\pi}{2})\cdot\mathbb{R}_{z}(\alpha_{0} - \frac{\pi}{2})\cdot\mathbf{V} \notag \\
	\mu_{\alpha} &= -V_{sky}(x)\sin(\alpha) + V_{sky}(y)\cos(\alpha) \notag \\
	\mu_{\delta} &= V_{sky}(z)\cos(\delta) -(V_{sky}(x)\cos(\alpha) + V_{sky}(y)\sin(\alpha))\sin(\delta)  \notag
\end{flalign}
This projection sets up a correlation between proper motion and in-plane velocity $v_{phi},v_{rot}$:
\begin{flalign}
    \mu_{\alpha} =&  P_1 v_{\phi} + P_2 v_{r} + V\alpha_{bulk}(\alpha,\delta) \notag \\
    \mu_{\delta} =&  P_3 v_{\phi} + P_4 v_{r} + V\delta_{bulk}(\alpha,\delta)\notag \\
    P_1 =& \cos(i)\cos(\phi)(\cos({\alpha_0})(\sin({\delta_0})\cos(\theta)\sin({\alpha}) -  \sin(\theta)\cos({\alpha})) - \sin({\alpha_0})(\sin({\delta_0})\cos(\theta)\cos({\alpha}) + \notag \\
        & \sin(\theta)\sin({\alpha}))) - \cos({\delta_0})\sin(i)\cos(\phi)\sin({\alpha} - {\alpha_0}) - \sin(\phi)(\sin({\delta_0})\sin(\theta)\sin({\alpha})\cos({\alpha_0})- \notag \\ 
        & \sin({\delta_0})\sin(\theta)\cos({\alpha})\sin({\alpha_0}) +  \cos(\theta)\cos({\alpha})\cos({\alpha_0})+\cos(\theta)\sin({\alpha})\sin({\alpha_0})) \notag \\
    P_2 =& (\sin (\phi ) + \cos (\phi )) (\cos (i) (\cos ({\alpha_0}) (\sin ({\delta_0}) \cos (\theta ) \sin ({\alpha}) -\notag \\
        & \sin (\theta ) \cos ({\alpha}))-\sin ({\alpha_0}) (\sin ({\delta_0}) \cos (\theta ) \cos ({\alpha}) + \sin (\theta ) \sin ({\alpha})))-\cos ({\delta_0}) \sin (i) \sin ({\alpha}-{\alpha_0}))\notag \\
    P_3 =& \cos (\phi ) (\cos ({\delta}) (\cos ({\delta_0})  \cos (\theta ) \cos (i)+ \sin ({\delta_0}) \sin (i))+ \sin ({\delta}) (\cos (i) (\sin ({\delta_0}) \cos (\theta ) \cos ({\alpha}) \cos ({\alpha_0})+ \notag \\
        & \sin ({\delta_0}) \cos (\theta ) \sin ({\alpha}) \sin ({\alpha_0})+  \sin (\theta ) \sin ({\alpha}) \cos ({\alpha_0})-\sin (\theta ) \cos ({\alpha}) \sin ({\alpha_0}))- \notag \\
        & \cos ({\delta_0}) \sin (i) \cos ({\alpha}-{\alpha_0})))- \sin (\phi ) (\sin (\theta ) (\sin ({\delta}) \sin ({\delta_0}) \sin ({\alpha}) \sin ({\alpha_0})+\notag \\
        & \cos ({\delta}) \cos ({\delta_0}))+\sin ({\delta}) \cos ({\alpha_0}) (\sin ({\delta_0}) \sin (\theta ) \cos ({\alpha})-  \cos (\theta ) \sin ({\alpha}))+\sin ({\delta}) \cos (\theta ) \cos ({\alpha}) \sin ({\alpha_0}))\notag \\
    P_4 =& \sin (\phi ) (\cos ({\delta}) (\cos ({\delta_0}) \cos (\theta ) \cos (i)+ \sin ({\delta_0}) \sin (i))+ \sin ({\delta}) (\cos (i) (\sin ({\delta_0}) \cos (\theta ) \cos ({\alpha}) \cos ({\alpha_0})+ \notag \\
        & \sin ({\delta_0}) \cos (\theta ) \sin ({\alpha}) \sin ({\alpha_0})+ \sin (\theta ) \sin ({\alpha}) \cos ({\alpha_0})-\sin (\theta ) \cos ({\alpha}) \sin ({\alpha_0})) -\notag \\
        & \cos ({\delta_0}) \sin (i) \cos ({\alpha}-{\alpha_0})))+ \cos (\phi ) (\sin (\theta ) (\sin ({\delta}) \sin ({\delta_0}) \sin ({\alpha}) \sin ({\alpha_0}) +\notag \\
        & \cos ({\delta}) \cos ({\delta_0}))+\sin ({\delta}) \cos ({\alpha_0}) (\sin ({\delta_0}) \sin (\theta ) \cos ({\alpha})- \cos (\theta ) \sin ({\alpha}))+\sin ({\delta}) \cos (\theta ) \cos ({\alpha}) \sin ({\alpha_0})) \notag 
\end{flalign}
And on the other hand, we have:
\begin{flalign}
    v_{\phi} =& A_1 \mu_{\alpha} + A_2 \mu_{\delta}   \notag \\ 
    v_{r} =& B_1 \mu_{\alpha} + B_2 \mu_{\delta}  \notag \\
    A_1 =& \frac{-P_4}{P_2P_3 - P_1P_4},\ A_2 = \frac{P_2}{P_2P_3 - P_1P_4} \notag \\
    B_1 =& \frac{P_3}{P_2P_3 - P_1P_4},\ B_2 = \frac{-P_1}{P_2P_3 - P_1P_4} \notag 
\end{flalign}
Since this is a linear transformation of a two dimensional Gaussian probability distribution, we expect the probability distribution in proper motion space is also a Gaussian distribution. The standard deviation and mean values are:
\begin{flalign}
     \sigma_{\alpha,Model}^2 &= P_1^2\sigma_{\phi}^2 + P_2^2\sigma_{r}^2 \notag \\
     \sigma_{\delta,Model}^2 &= P_3^2\sigma_{\phi}^2 + P_4^2\sigma_{r}^2\notag \\
     \mu_{\alpha,Model} &= P_1 V_{\phi}(r) + V\alpha_{bulk}(\alpha,\delta) \notag \\
     \mu_{\delta,Model} &= P_3 V_{\phi}(r) + V\alpha_{bulk}(\alpha,\delta) \notag 
\end{flalign}
We incorporate the data uncertainty as:
\begin{flalign}
     \sigma_{\alpha}^2 &= \sigma_{\alpha,Model}^2 + \sigma_{\alpha,Data}^2 \notag \\
     \sigma_{\delta}^2 &= \sigma_{\delta,Model}^2 + \sigma_{\delta,Data}^2 \notag
\end{flalign}
Then the correlation is
\begin{equation}
    \rho = \frac{A_1B_1 \sigma_{\phi}^2 + A_2B_2 \sigma_{r}^2}{\sigma_{\alpha}\sigma_{\delta}} \notag
\end{equation}
The probability distribution in proper motion space:
\begin{flalign}
    p(\mu_{\alpha},\mu_{\delta}) =& \frac{1}{2\pi \sigma_{\alpha}\sigma_{\delta}\sqrt{1-\rho^2}}\exp\left[-\frac{1}{(1-\rho^2)}\left(\frac{(\mu_{\alpha} - \mu_{\alpha,Model})^2}{2\sigma_{\alpha}^2} + \right.\right. \left.\left. \frac{(\mu_{\delta} - \mu_{\delta,Model})^2}{2\sigma_{\delta}^2} - \rho\frac{(\mu_{\alpha} - \mu_{\alpha,Model})(\mu_{\delta} - \mu_{\delta,Model})}{\sigma_{\alpha}\sigma_{\delta}}\right)\right] \notag
\end{flalign}

\section{MCMC sampling results for RGB and young stars}

\begin{figure*}
    \includegraphics[width=\textwidth]{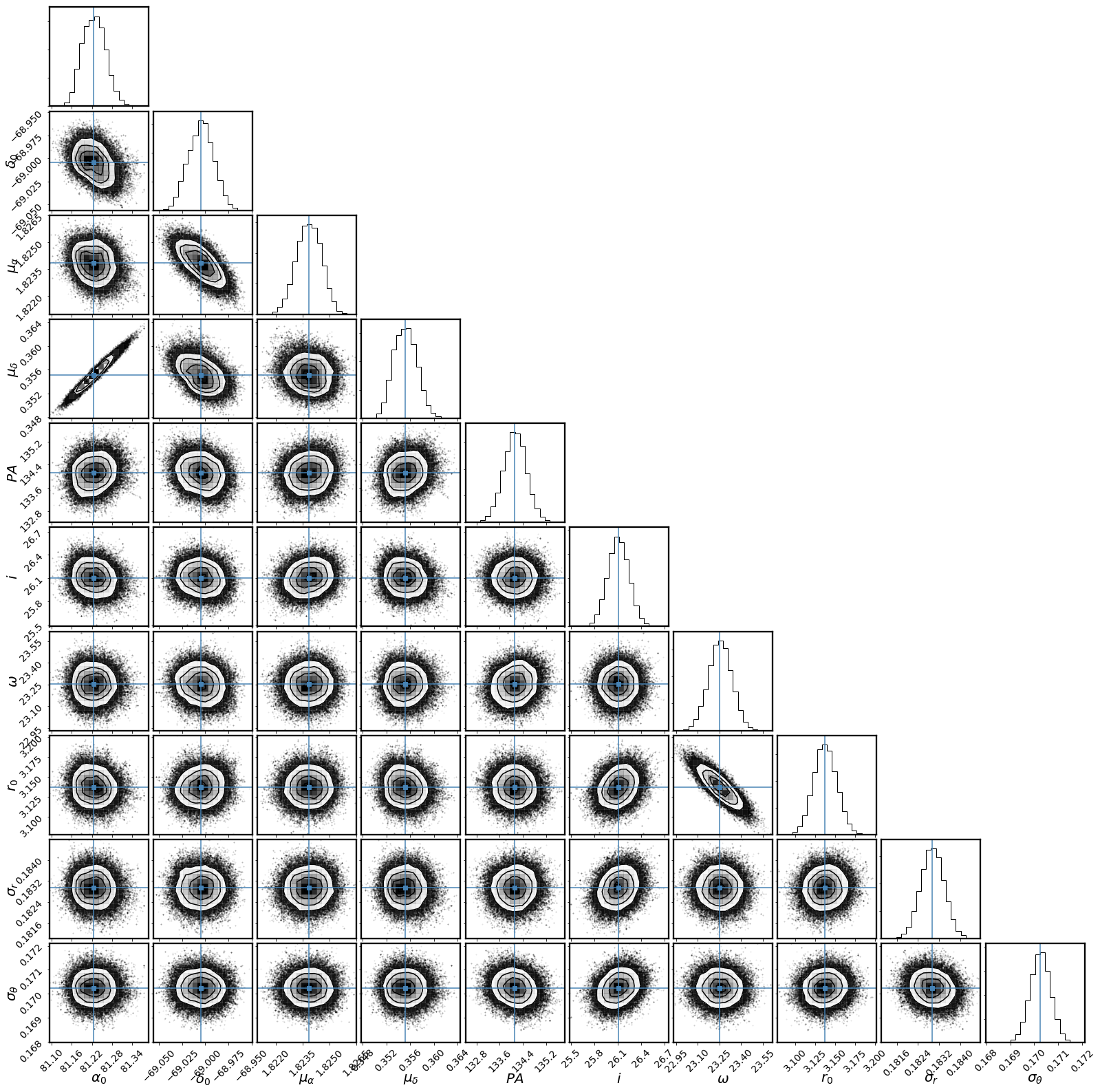}
        \caption{Corner plot summary of the MCMC sampling result for the RGB stars. The sample size for this population is much larger than for the Carbon Stars, but the degree of contamination is also greater.}
        \label{Fig:mcmc_configure_fit_P4}
\end{figure*}

\begin{figure*}
    \includegraphics[width=\textwidth]{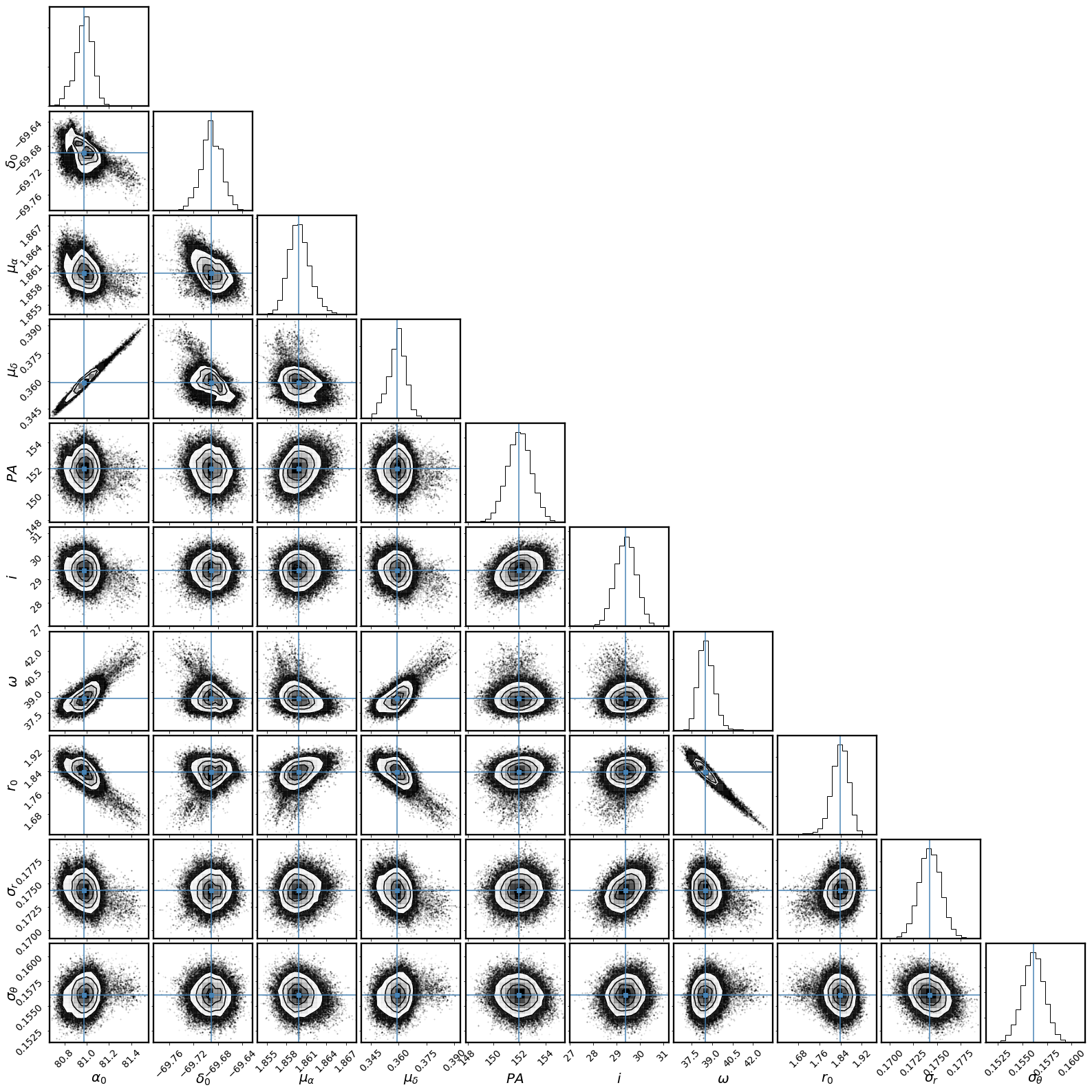}
        \caption{Corner plot summary of the MCMC sampling result for the young MS stars. There are similar correlations evident as in Fig.\ref{Fig:mcmc_configure_fit}.}
        \label{Fig:mcmc_configure_fit_P1}
\end{figure*}


\bsp	
\label{lastpage}
\end{document}